\documentclass[12pt]{article}
\usepackage{psfig,epsfig}
\begin{document}

\begin{large}
  \title{\bf Nuclear {\mbox{\boldmath{$\gamma$}}}-radiation as a
    Signature of Ultra Peripheral Ion Collisions at LHC energies}
  \author{
    Yu.V.Kharlov\thanks{Institute for High Energy Physics,
      142281 Protvino, Russia},
    V.L.Korotkikh\thanks{Scobeltsyn Institute of Nuclear Physics,
      Moscow State University, 119992 Moscow, Russia}}
  \date{}
  \maketitle
  
\begin{abstract}
\bigskip 

\begin{normalsize}
  We study the peripheral ion collisions at LHC energies in
  which a nucleus is excited to the discrete state and then emits
  $\gamma$-rays.  Large nuclear Lorenz factor allows to observe the
  high energy photons up to a few ten GeV and in the region of angles
  of a few hundred micro-radians around the beam direction. These
  photons can be used for tagging the events with particle production
  in the central rapidity region in the ultra-peripheral collisions. For
  that it is necessary to have an electromagnetic detector in front of
  the zero degree calorimeter in the LHC experiments.
\end{normalsize}
\end{abstract} 

\section*{Introduction}
There are several reviews devoted to the coherent $\gamma \gamma$ and
$\gamma g$ interactions in the very peripheral collisions at
relativistic ion colludes
(\cite{Bertulani88},\cite{Baur98},\cite{Baur}). The advantage of
relativistic heavy ion colliders is that the effective photon luminosity for
two-photon physics is of orders of magnitude higher than the one at
available the $e^+e^-$ machines. There are many suggestions to use the
electromagnetic  interactions of nuclei to study production of
meson resonances, Higgs boson, Radion scalar or exotic mesons.
These interactions allow also to study fermion, vector meson or
boson pair production, as well as to investigate a few new physic
regions (see list in \cite{Baur}). The $\gamma g$ interactions will
open a new page of nuclear physics such as a study of nuclear gluon
distribution. It is also important for a knowledge of the details of
medium effects in nuclear matter at the formation of quark-gluon
plasma \cite{Goncalves}. These effects may be studied by
photo-production of heavy quarks in virtual photon-gluon interactions
(\cite{Greiner},\cite{Klein},\cite{Goncalves}).

For these investigations it is necessary to select the processes with
large impact parameters $b$ of colliding nuclei, $b>(R_1+R_2)$, to
exclude background from strong interactions. Note, that some
processes, like $\gamma \gamma$-fusion to Higgs boson or Radion
scalar, are free from any problems caused by strong interactions of
the initial state \cite{Lietti}.  Therefore we need an efficient
trigger to distinguish $\gamma \gamma$ and $\gamma g$ interactions
from others. G.Baur et al. \cite{Baur99} suggested to measure the intact
nuclei after the interaction. Evidently this is impossible in the LHC
experiments since the nuclei fly into the beam pipe.

It is interesting to consider a $\gamma$ rays emitted by the 
relativistic nuclei at LHC energies. Such kind of process was 
used for the possible 
explanation of the high energy ($E_\gamma \geq 10^{12}$ eV) cosmic photon
spectrum~\cite{Balashov}.

It was suggested to measure a nuclear $\gamma $ radiation after the
excitation of discrete nuclear level in our work \cite{Korotkikh}.
These secondary photons have the energy of a few GeV and the narrow
angular distribution near the beam direction due to a large Lorentz
boost.  The angular width is enough to register them in the
electromagnetic zero-degree detectors of the future LHC experiments
CMS or ALICE. A nucleus saves its $Z$ and $A$ in this process. So we
have a clear electromagnetic interaction of nuclei at any impact
parameter. The nuclear $\gamma $ radiation may be used as
``event-by-event'' criteria for such kind of collisions.

We have considered \cite{Korotkikh} only the process $A+A \to
A^*+A+e^+e^-, A^* \to A+\gamma '$, where a nucleus is excited by
electron (positron) ~ $e^\pm +A \to e^{\pm'}+A^*$.
 Now we calculate the production process of some system $X_f$ 
in $\gamma \gamma$ fusion with simultaneous excitation of discrete
nuclear level.

In this work we consider the processes
$$
^{16}{\mbox O} + ^{16}{\mbox O} \to  ^{16}{\mbox O} + ^{16}{\mbox 
  O}^*(2^+,6.92~{\mbox {MeV}}) +X_f,  ^{16}{\mbox O}^* \to^
{16}{\mbox O}+\gamma,
$$
$$
^{208}{\mbox {Pb}} +^{208}{\mbox {Pb}} \to^{208}{\mbox {Pb}}
+^{208}{\mbox {Pb}}^*(3^-,2.62~{\mbox {MeV}}) +X_f, ^{208}{\mbox {Pb}}^* 
\to^{208}{\mbox {Pb}}+\gamma,
$$
where the  $^{16}$O and $^{208}$Pb were taken since they are
the lightest and heaviest ions in
 the ion list of the LHC program. The trigger requirements
will include a signal in the central rapidity region of particles from
$X_f$ decay, a signal of photons in the electromagnetic detector in
front of the zero degree calorimeter and a veto signal of neutrons in
ZDC. We suggest to use the veto signal of neutrons in order to avoid
the processes with the nuclear decay into nucleon fragments.

The formalism of the considered process is presented in the section 1.
The nuclear form factors are calculated in the section 2. The angular
and energy distributions of secondary photons are in the section 3.
The cross sections of $\eta_c(2.979~\mbox{GeV})$ production are
presented in the part 6 with and without nuclear excitation. The
section 6 is our conclusion.
 
\section{Formulae of nuclear excitation cross-section and photon
  luminosity in peripheral interactions} 

Let us consider the peripheral ion collision
\begin{equation}
\label{1}
A_1 + A_2 \to A_1^* (\lambda
^P,E_0) + A_2 + X_f,
\end{equation}
where $X_f$ is the produced system in $\gamma^*\gamma^*$ fusion and
$A_1^*$ is an excited nucleus in a discrete nuclear level with
spin-parity $\lambda^P$ and energy $E_0$ (see Fig.\ref{fig0_gam}).
Here the nucleus $A_1$ and $A_2$ have equal mass $A$ and charge $Z$,
the only nucleus $A_1$ is excited.  We suppose that the particles of
$X_f$ decay can be registered in the central rapidity region. The
nuclear $\gamma~'$ radiation ($A_1^*\to A_1+\gamma~'$) will be
measured in the forward detectors such as ZDC.
\begin{figure}[ht]
  \centerline{
    \epsfig{file=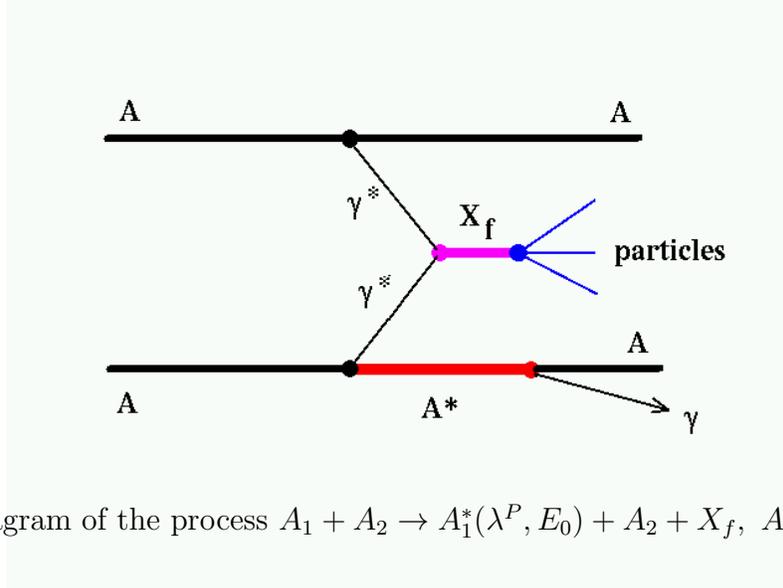,width=100mm,angle=0,bb=0 100 611 460}
  }
  \caption{Diagram of the process $A_1 + A_2 \to A_1^* (\lambda^P,E_0)
    + A_2 + X_f, ~A_1^*\to A_1+\gamma $.}
  \label{fig0_gam}  
\end{figure}

\bigskip

We use the quantum mechanical plane wave formalism (\cite{Budnev},\cite{Baur})
and the derivation of
 the equivalent photon approximation. It allows us to introduce the elastic and inelastic nuclear form factors for the process (\ref{1}).  
We take the formulae (19) and (21) in \cite{Baur} :
\begin{equation}
\label{2}
d\sigma_{A_1A_2 \to A_1^*A_2X_f} = \int \frac{dw_1}{w_1}\int \frac{dw_2}{w_2}
n_1(w_1)n_2(w_2)d\sigma_{\gamma\gamma\to X_f}(w_1,w_2),
\end{equation}
\begin{equation}
\label{3}
n_i(w_i)=\frac{\alpha}{\pi^2}\int d^2q_{i\perp} \int
d\nu_i\frac{1}{(q_i^2)^2}\left[2\frac{w_i^2m_i^2}{P_i^2}W_{i,1} +
q_{i\perp}^2W_{i,2}\right],
\end{equation}
where $W_{i,1}$ and $W_{i,2}$ are the Lorentz scalar functions. All kinematic
variables are the same as in \cite{Baur}. 

For ``elastic'' photon process $A_1A_2 \to A_1A_2X_f$ we have 
\begin{equation}
\label{4}
W_1=0,~W_2(\nu,q^2)=Z^2F_{el}^2(-q^2)\delta(\nu+q^2/2m)
\end{equation}
\begin{equation}
\label{5}
n(w) =
\frac{Z^2\alpha}{\pi^2}\int  d^2q_\perp\frac{q_\perp^2}{(q^2)^2}F_{el}^2(-q^2),
\end{equation}
where $F_{el}(q)$ is the nuclear form-factor with $F_{el}(0)=1$. 

For the excitation of nucleus to a discrete state with a spin $\lambda$
and an energy $E_0$ (``inelastic'' photon process $A_1A_2 \to A_1^*
(\lambda^P,E_0) A_2X_f$)
\begin{equation}
  \label{6}
  \begin{array}{rcl}
    W_{1,2}(\nu,q^2) & = &\hat{W}_{1,2}(q^2)\delta(\nu-E_0), \\
    -q^2 & = & \displaystyle{\frac{w^2}{\gamma^2}+2\frac{wE_0}{\gamma} + \frac{E_0^2}{\gamma^2} +
      q_\perp^2} = q_L^2(w) + q_\perp^2, \\
    \hat{W}_1 & = & 2\pi[|T^e|^2 + |T^m|^2], \\
    \hat{W}_2 & = & \displaystyle{2\pi\frac{q^4}{(E_0^2 - q^2)^2} \left[2|M^c|^2 -
        \frac{E_0^2-q^2}{q^2}(|T^e|^2 + |T^m|^2)\right]}. 
  \end{array}
\end{equation}
See notations again in \cite{Baur}. 

We neglect the transverse electric $T^e$ and transverse magnetic
$T^m$ matrix elements comparing with the Coulomb one $M^c \equiv M_\lambda$
for $0^+ \to \lambda^P$ nuclear transitions. Then for the inelastic photon
process with a nuclear discrete state excitation we get
\begin{equation}
  \label{7}
  n_1^{(\lambda)}(w) = \frac{4\alpha}{\pi}\int d^2q_\perp
  \frac{q_\perp^2}{(E_0^2-q^2)^2}|M_\lambda(q)|^2,
\end{equation}
where $M_\lambda(q)$ is the inelastic nuclear form-factor.

The equivalent photon number (\ref{7}) can be represented as the
function of $q_\perp$ for inelastic photon emission:
\begin{eqnarray}
\label{19}
\frac{dN_1^{(\lambda)}}{dq_\perp^2}(w_1,q_\perp) & = & \frac{4\alpha}{\pi}
\frac{q_\perp^2}{(E_0^2-q^2)^2}|M_\lambda(-q^2)|^2 = \nonumber \\
\label{20}
& = &
\frac{4\alpha}{\pi}\Bigg|\frac{q_\perp}{(E_0^2-q^2)}M_\lambda(-q^2) 
e^{i\varphi_\perp}\Bigg|^2,
\end{eqnarray}
where $q_\perp e^{i\varphi_\perp}= \vec{q_\perp}$ (see \cite{Baur2}).

Let us do the inverse transformation to the impact parameter $b$ presentation 
\begin{equation}
\label{21}
f(\vec{b}) = \frac{1}{2\pi} \int d^2q_\perp e^{-i\vec{q}_\perp\vec{b}}f(\vec{q}_\perp).
\end{equation} 

For the function under the module in equation (\ref{20}) we get 
\begin{eqnarray}
\label{22}
f(\vec{b}) & = & \frac{1}{2\pi}\int d^2q_\perp \frac{q_\perp}{(E_0^2-q^2)}
M_\lambda(-q^2)e^{i\varphi_\perp}\cdot e^{-i\vec{q}_\perp\vec{b}} = \nonumber \\
\label{23}
& = & i\int dq_\perp \frac{q_\perp^2}{(E_0^2-q^2)}M_\lambda(-q^2)\cdot J_1(q_\perp b) =
\nonumber \\
\label{24}
& = & \frac{i}{b} \int du \frac{u^2}{u^2+(E_0^2+q_L^2)~b^2}
M_\lambda\left(-\frac{x^2+u^2}{b^2}\right) J_1(u).
\end{eqnarray} 

If we take  $M_{el}$ instead of the inelastic $M_\lambda$ as
\begin{equation}
\label{24a}
|M_{el}(-q^2)|^2 = \frac{Z^2}{4\pi}F^2_{el}(-q^2)
\end{equation} 
we get a well-known formula for elastic photon process  
(see (\ref{4}) in \cite{Baur2}) where $F_{el}(0)=1$:
\begin{equation}
\label{25}
N_2^{(el)}(w,b) = \frac{Z^2\alpha}{\pi^2}\frac{1}{b^2} \Bigg| \int du
\frac{u^2}{x^2+u^2} J_1(u) F_{el}[-(x^2+u^2)/b^2]\Bigg|^2,
\end{equation}
Here $x=q_L b=wb/\gamma_A $ and $u=q_\perp b$. 
For a point charge, $F_{el}(q)\equiv 1$, we readily obtain
\begin{equation}
\label{25e}
N_2^{(el)}(w,b) = \frac{Z^2\alpha}{\pi^2}\frac{1}{b^2}~x^2~K_1^2(x) ,
\end{equation}
in agreement with \cite{Baur} at very large $\gamma_A$.

We write the form factors of elastic and inelastic nuclear process 
  in the same forms:
\begin{eqnarray}
\label{rho5}
\mbox{$\cal F$}^2 _{\lambda}(q)& =& \frac{1}{4\pi e^2Z^2} F^2_{\lambda}(q) \\
\label{rho6}
 F^2_{0}(q) & =& \Bigg | 4\pi \frac{1}{q} \int \sin(qr)
 \rho_0(r)rdr \Bigg |^2  _{q\to 0} \to 1,  \\ 
F^2_{\lambda}(q) & =& (2\lambda+1)\Bigg | 4\pi  \int j_\lambda(qr)
 \rho_\lambda(r,Z)r^2dr \Bigg |^2 _{q\to 0} \to\\
&\to& \frac {(4 \pi)^2 B(E\lambda)}{e^2Z^2[(2\lambda +1)!!]^2}~q{2\lambda},
\end{eqnarray}
where $ \rho_\lambda(r,Z)$ is an  nuclear transition density and
$B(E_0\lambda)$ is the reduced transition
probability .

Then for the matrix elements $M_\lambda$ we get in the limit $q\to 0 $ 
\begin{equation}
  \label{8a}
  |M_{el}(-q^2)|^2 =
  \left.\left(\frac{Z^2}{4\pi}\right) F_{el}^2(q)~\right|_{q\to 0}
  \to\frac{Z^2}{4\pi}
\end{equation}
\begin{equation}
  \label{9}
  \left.|M_\lambda(-q^2)|^2 =
\left(\frac{Z^2}{4\pi}\right) F_{\lambda}^2(q)
~\right|_{q\to 0} \to
  \left(\frac{Z^2}{4\pi}\right) \frac{(4\pi)^2 B(E_0\lambda)}{e^2Z^2[(2\lambda+1)!!]^2}q^{2\lambda}.
\end{equation} 

The effective photon number for inelastic process with nuclear
transition $0 \to \lambda$ will be
\begin{equation}
  \label{26}
  N_1^{(\lambda)}(w,b) =
  \frac{Z^2\alpha}{\pi^2}\frac{1}{b^2} \Bigg| \int\limits_0^\infty du
  \frac{u^2}{x_{in}^2+u^2} J_1(u) F_\lambda[-(x_{in}^2+u^2)/b^2]\Bigg|^2,
\end{equation}
as the generalization of (\ref{25}). Here
$x_{in}^2=(E_0^2+\frac{w^2}{\gamma^2}+2\frac{wE_0}{\gamma} +
\frac{E_0^2}{\gamma^2})~b^2$.

We take the inelastic form-factor from inelastic electron scattering 
off nuclei. A good parameterization of inelastic form-factor 
is 
\begin{equation}
\label{10}
F^2_\lambda(q) = 4\pi \beta_\lambda^2 j_\lambda^2(qR)e^{\displaystyle -q^2g^2}
\end{equation}
in the Helm's model \cite{Helm}. The squared transition radius is equal to 
$R_\lambda^2=R^2+(2\lambda+3)g^2$, where $g$ is a size of a nuclear diffusion side.

The reduced transition probability in this case
is equal to 
\begin{equation}
\label{10a} 
B(E_0\lambda) = \frac{\beta_\lambda^2}{4\pi} Z^2e^2R^{2\lambda}. 
\end{equation}

So, the formulae for the process (\ref{1}) are 
\begin{eqnarray}
\label{11}
d\sigma_{A_1A_2 \to A_1^*A_2X_f} & = & \int\frac{dw_1}{w_1}\int \frac{dw_2}{w_2}
n_1^{(\lambda)}(w_1)n_2(w_2)d\sigma_{\gamma\gamma \to X_f}(w_1,w_2); \\
\label{12}
n_1^{(\lambda)}(w_1) & = & \frac{Z^2\alpha}{\pi^2}\int d^2q_\perp \frac{q_\perp^2}{(E_0^2 -q_{in}^2)^2} 
|F_\lambda(-q_{in}^2)|^2; \\
\label{13}
-q_{in}^2 & = & \frac{w^2}{\gamma_A^2} + 2\frac{wE_0}{\gamma_A} +
\frac{E_0^2}{\gamma_A^2} + q_\perp^2; \\
\label{15}
n_2(w_2) & = & \frac{Z^2\alpha}{\pi^2} \int d^2q_\perp \frac{q_\perp^2}{q_{el}^4}
F_{el}^2(-q_{el}^2); \\
\label{16}
-q_{el}^2 & = & \left(\frac{w}{\gamma_A}\right)^2 + q_\perp^2.
\end{eqnarray} 

\noindent{The value $q_{in}^2$ is close to $q_{el}^2$ at a large
$\gamma_A$ factor at LHC energies.}

The effective two photon luminosity can be expressed as
\begin{equation} 
L(\omega_1,\omega_2)= 
2 \pi \int\limits_{R_1}^\infty b_1 db_1  \int\limits_{R_2}^\infty b_2 db_2  
\int\limits_0^{2 \pi} d \phi 
N_1^{(\lambda)}(\omega_1,b_1) N_2^{(el)}(\omega_2,b_2) \Theta(B^2), 
\label{aaspectr} 
\end{equation} 
where $R_1$ and $R_2$ are the nuclear radii, $\Theta(B^2)$ is the step
function and $B^2=b_1^2+b_2^2-2 b_1 b_2 \cos{\phi}-(R_1+R_2)^2$
\cite{Baur}.  Then the final cross-section is
\begin{equation} 
\label{33}
\sigma_{A_1A_2 \to A_1^*A_2X_f}=\int {d\omega_1\over \omega_1}\int {d\omega_2\over \omega_2} 
L(\omega_1,\omega_2)~\sigma_{\gamma\gamma \to X_f}(w_1,w_2) 
\end{equation}

\section{Nuclear levels and form-factors}

The elastic form factor of a light nucleus is 
\begin{equation} 
  \label{17}
  F_{el}(q^2)  =
  \exp\left(\displaystyle -\frac{\langle r^2\rangle}{6}~q^2\right)
\end{equation}
with $~\sqrt{\langle r^2 \rangle} = 2.73~{\mbox{fm}}$ for the nucleus
$^{16}$O.  For a heavy nucleus we take a modified Fermi nuclear density
\cite{Eldyshev}
\begin{eqnarray}  
\label{17a}
\rho(r) & = &\rho_0~\left\{
\frac{1}{1+exp\frac{-r-R}{g}} +
\frac{1}{1+exp\frac{r-R}{g}} -1
\right\}\\
 & = &\rho_0~
\frac{sh(R/g)}{ch(R/g)+ch(r/g)},\\
\rho_0 & = & \frac{3}{4\pi R^3}\left\{1+\left(\frac{\pi g}{R}\right)^2 
\right\}^{-1} 
\end{eqnarray} 
with the parameters for $^{208}{\mbox{Pb}}$ are equal to
$R  =  6.69~{\mbox{fm}}, g = 0.545~{\mbox{fm}}.$ 
Such form of density  is close to the usual Fermi density at $g\ll R$
\begin{equation} 
\label{17b}
\rho_F(r)= \rho_0~\frac{1}{1+\exp\frac{r-R}{g}}
\end{equation} 
and allows us to calculate analytically the elastic form factor
\begin{equation} 
\label{17c}
F_{el}(q) =\frac{4\pi^2 R g\rho_0}{q~{\rm sh}(\pi g q)}
\left\{\frac{\pi g}{R}~\sin(q R)~{\rm cth}(\pi g q)-\cos(q R) \right\}. 
\end{equation} 

There are a few discrete levels of $^{16}$O below $\alpha$, $p$ and
$n$ thresholds $E_{th}(\alpha)=7.16$ MeV, $E_{th}(p)=7.16$ MeV,
$E_{th}(n)=7.16$ MeV \cite{Endt}. The level $2^+$ at $E_0=6.92$ MeV is
the strongest excited one in the electron scattering.

The parameters from the inelastic electron scattering fit on  $^{16}$O
with excitation of $2^+$ level ($E_0=6.92$ MeV) of $^{16}$O are \cite{Gulk}:
$$\beta_2 = 0.30,~~R = 2.98~{\mbox{fm}},~~g =
0.93~{\mbox{fm}}.$$
They correspond to
\begin{equation}
\label{10g}
B(E_02) = (36.1\pm3.4)e^2~{\mbox{fm}}^4.
\end{equation}

There are more than 70 discrete levels of  $^{208}$Pb \cite{NDS} below the 
neutron threshold $E_{th}(n)=7.367$ MeV. About 30\% of the levels decay to 
the first $3^-$ level of $^{208}$Pb at  $E_0=2.615$ MeV. This level is well 
studied experimentally \cite{Goutte} and has a large excited cross-section.

The reduced transition probability from the fit of inelastic electron scattering on $^{208}$Pb
with excitation of the  $3^-$ level  is  \cite{Goutte}:
\begin{equation}
\label{10h}
B(E_03) = (6.12~10^5\pm~2.2\%)e^2~{\mbox{fm}}^6.
\end{equation}

We calculate the parameter $\beta_3$, using this $B(E_03)$, and take $R$ and  $g$ from the density 
of the $^{208}Pb$ ground  state:
$$\beta_3 = 0.113,~~R = 6.69~{\mbox{fm}},~~g =
0.545~{\mbox{fm}}.$$

Note that there are many levels higher than $E_0=2.615$ MeV which decay to the 
first level of $^{208}$Pb. This fact increases the event rate of the
process (\ref{1}), but we don't know cross-section excitation of these levels.

The elastic form factor (\ref{17}) of $^{16}$O and inelastic
form-factor $^{16}{\mbox O}~(2^+, 6.92$ MeV) (\ref{10}), corresponding
to the electron scattering data, are shown in Fig.\ref{fig1_gam}.
The same for a nucleus $^{208}$Pb and the exited state
$^{208}{\mbox{Pb}} ~(3^-, 2.64$~MeV) are shown in Fig.\ref{fig2_gam}.
\begin{figure}[htp]
  \centerline{
    \epsfig{file=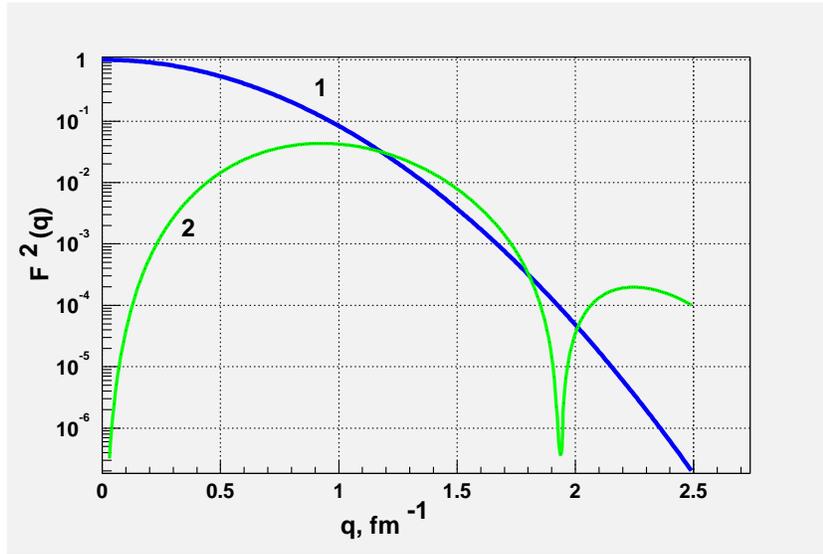,width=110mm,angle=0}
  }
  \caption{The elastic  form-factor of $^{16}$O  (1) and the inelastic
    form-factor of $^{16}{\mbox O}~(2^+, 6.92$ MeV) (2) from the
    electron scattering.}
  \label{fig1_gam}  
\end{figure}

\begin{figure}[htp]
  \centerline{
    \epsfig{file=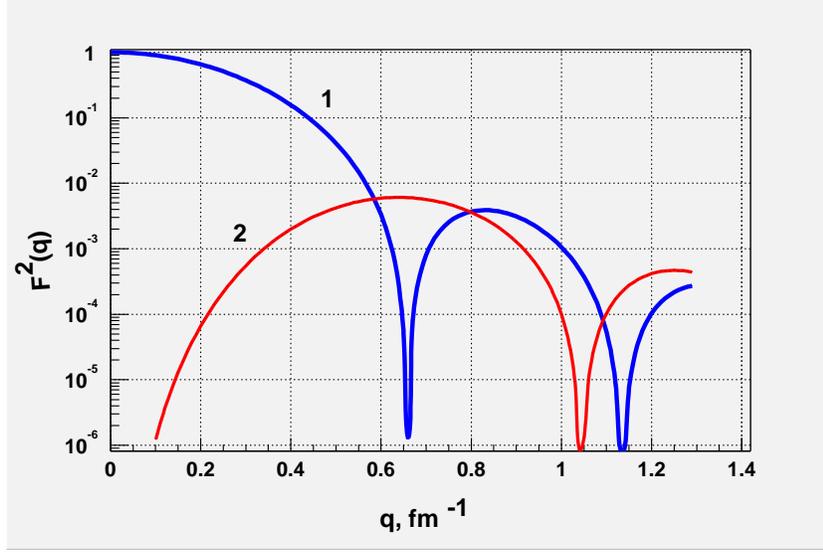,width=110mm,angle=0}
  }
  \caption{The elastic  form-factor of $^{208}$Pb  (1) and the
    inelastic  form-factor of $^{208}{\mbox{Pb}}~(3^-, 2.615$ MeV)
    (2).}
  \label{fig2_gam}  
\end{figure}

The squared inelastic form-factor is less than the elastic form-factor
by more then two orders at small $q < q_0$ ($q_0 = 1~{\mbox{fm}}^{-1}$
for $^{16}$O and $q_0 = 0.6~{\mbox{fm}}^{-1}$ for $^{208}$Pb).  In the
region of $q \simeq q_0$ they are comparable. The region of large $q >
q_0$ will give contribution for the small impact parameter $b$.  We
are able to calculate the photon luminosity (\ref{aaspectr}) for all
regions of $b$ to get the maximum electromagnetic cross-section of
process we are interested in. Then it should be possible to compare
with experimental data in the condition of clear selection of such
process by the photon signal and the veto neutron or proton signal in ZDC.

\bigskip

\section{Angular and energy distributions of secondary nuclear photons}

We suppose that the nucleus $A_1^*$ in the process (\ref{1}) is
unpolarized.  Just now we don't know the relative excitation
probability of $|\lambda\mu>$ state of $A_1^*$, where $\mu$ is a
projection of spin $\lambda$. This assumption needs the study in
future.  So we use a formula (27) in our work \cite{Korotkikh} for the
angular distribution of secondary photons, which is valid for
isotropic photon distribution in the rest system of $A_1^*$.

If we know the cross-section of reaction (\ref{1}) calculated by the
equation (\ref{33}) then the angular and energy distribution of
photons are equal to:
\begin{equation}
\label{46}
\begin{array}{rcl}
{\displaystyle{\frac{d\sigma_{A^*}}{d\theta_\gamma}}} & = & {\displaystyle{\sigma_{A_1A_2 \to A_1^*A_2X}\cdot
\frac{2\gamma_A^2\sin\theta_\gamma}{(1+\gamma_{A_1^*}^2\tan^2\theta_\gamma)^2 
\cdot \cos^3\theta_\gamma}}}.
\end{array}
\end{equation}

The photon energy $E_\gamma$ and polar angle $\theta_\gamma$ in
laboratory system are defined as:
\begin{eqnarray}
\label{47}
E_\gamma & = & \gamma_{A_1^*}E_0(1 + \cos\theta_\gamma ')=2\gamma_{A_1^*}E_0/(1 + \gamma_{A_1^*}^2 \tan^2\theta_\gamma ), \\
\label{48}
\tan \theta_\gamma  & = & \frac{1}{\gamma_{A_1^*}}~\frac{\sin\theta_\gamma '}
{1+\cos\theta_\gamma '}, 
\end{eqnarray}
where $\theta_\gamma '$ and $\theta_\gamma $ are polar angles of
nuclear photon in the rest nuclear system and in the laboratory system
 with an axis $\vec z||\vec p_{A^*}$. Photon energy $E_\gamma$ dependence  on $\theta_\gamma$  are shown in
Fig.\ref{fig4gam}.

Our calculations 
with the help of TPHIC event generator \cite{Hencken}  show that a
deflection of the direction $\vec p_{A^*}$ from $\vec p_{beam}$ at LHC
energies in the reaction (\ref{1}) is very small at large $ \gamma_A$, $\langle\Delta \theta \rangle
\simeq 0.5 ~\mu$rad.

\begin{figure}[htp]
  \centerline{
    \epsfig{file=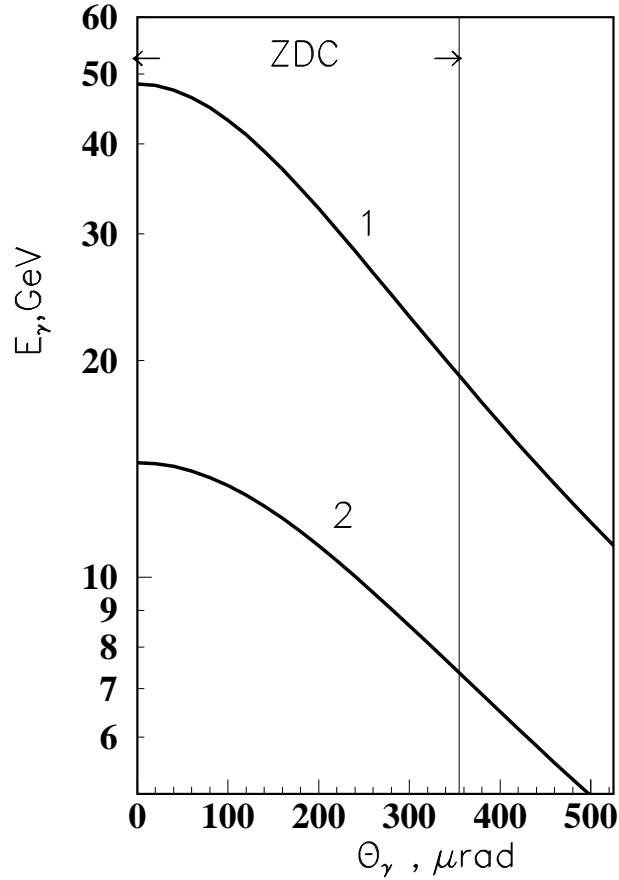 ,width=80mm,angle=0}
  }
  \caption{Nuclear photon energy as function of its polar angle in the
    laboratory system at LHC energies for two nuclei: $^{16}{\mbox
      0}~(2^+\to 0^+, 6.92$ MeV) (1)  and $^{208}{\mbox{Pb}}~(3^-\to
    0^+, 2.615$ MeV) (2). ZDC marks a region of Zero Degree
    Calorimeter in CMS.}
  \label{fig4gam}
\end{figure}

In the experiments CMS and ALICE, which are planned at LHC (CERN), the
Zero Degree Calorimeter (\cite{CMSZDC}, \cite{ALICE}) were suggested
for the registration of nuclear neurons after interaction of two ions.
We demonstrate a schematic figure of ZDC (CMS) at a distance
$L=140$~m in the plane transverse to the beam direction in
Fig.\ref{fig5gam}.  The CMS group plans to include also the
electromagnetic calorimeter in front of ZDC.
\begin{figure}[htp]
  \centerline{
    \epsfig{file=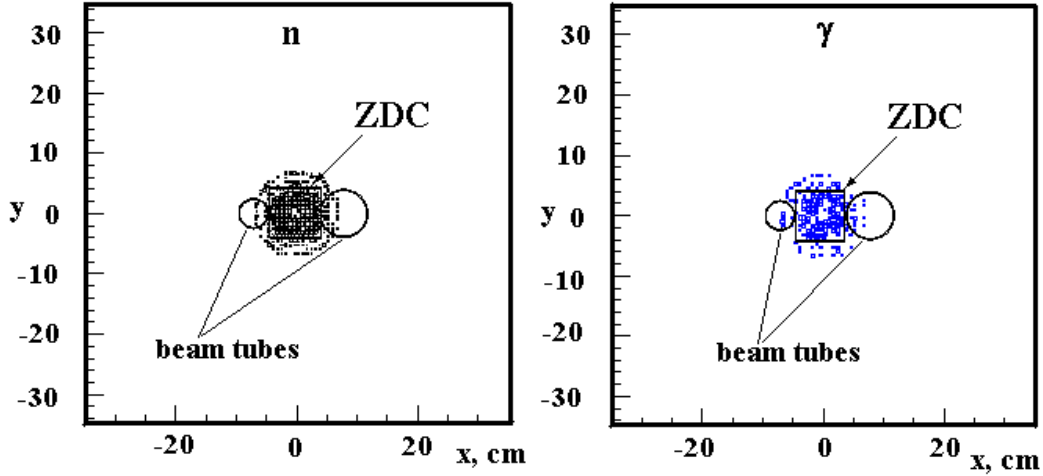,width=140mm,angle=0}
  }
  \caption{Transverse  ZDC plane. The points are the simulated
    hits of neutrons (left) and photons (right) from a work
    (\cite{ALICE}).}
  \label{fig5gam}
\end{figure}

As an example we demonstrate the angular distributions (\ref{46}) in
arbitrary units and energy dependence (\ref{47}) on the $(x,y)$
coordinates of ZDC (CMS) for two nuclei $^{16}$O and $^{208}$Pb in Fig.\ref{fig6gam}.
\begin{figure}[htp]
  \centerline{
    \epsfig{file=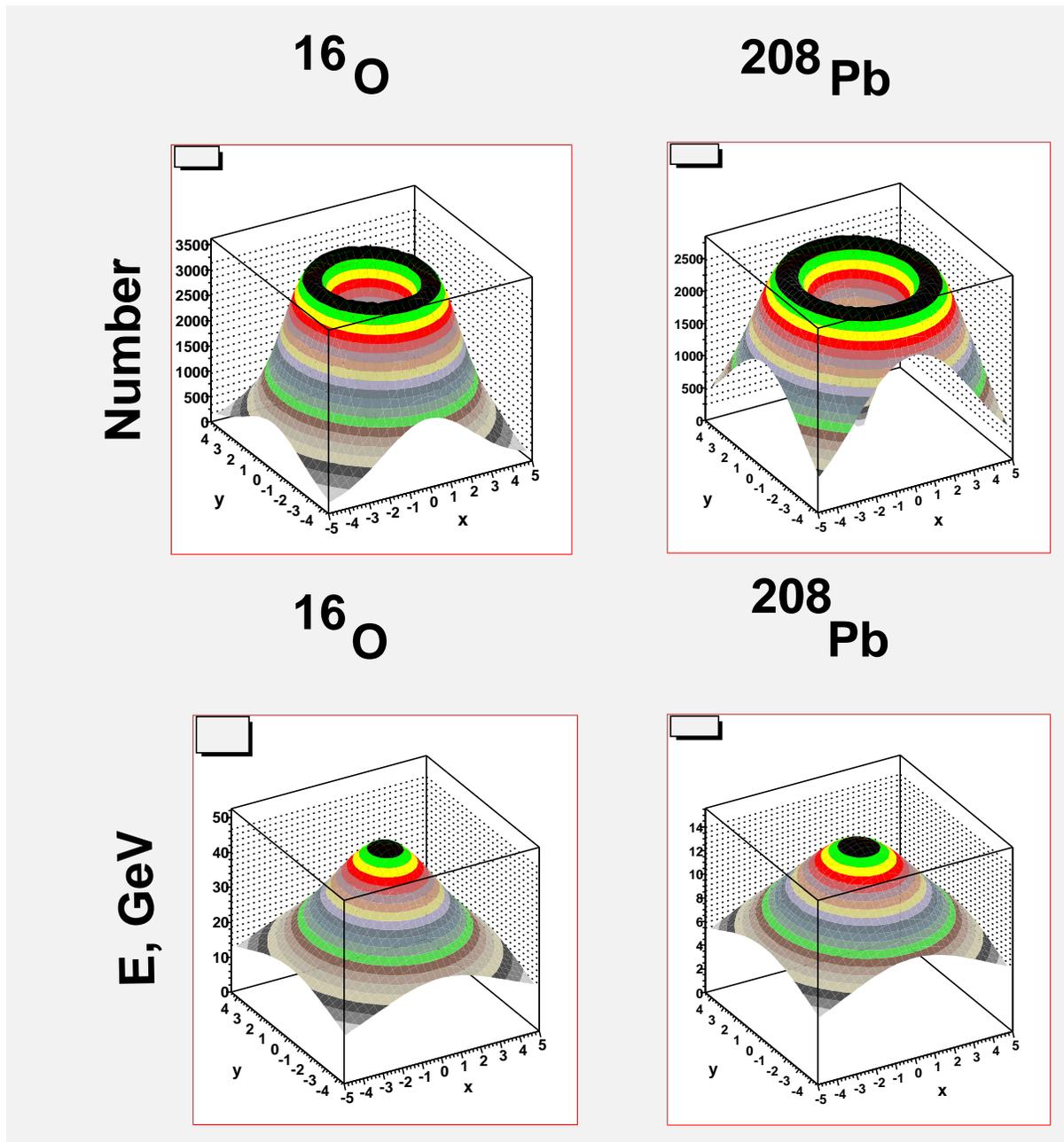 ,width=160mm,angle=0}
  }
  \caption{The photon angular distributions (upper raw) and the energy
    dependence (lower raw)  for
    $^{16}{\mbox O}^*(2^+, 6.92$ MeV) (left coulomb) and
    $^{208}{\mbox{Pb}}^*(3^-, 2.62$ MeV) (right coulomb) radiation
    decay in the laboratory system on the ZDC plane ($x,y$) at the
    distance 140 m from point interaction. $x$, cm is a horizontal and
    $y$, cm is a vertical axis.
    Photon energy interval in ZDC region is $19\div 48$ GeV for
    $^{16}{\mbox O}^*(2^+)$ and $7\div 14$ GeV for
    $^{208}{\mbox{Pb}}^*(3^-)$.}
  \label{fig6gam}  
\end{figure}
  The direction of the nucleus
$A_1^*$ coincides here the beam direction. A point $(x,y)=(0,0)$ is a
center of the ZDC plane.

\section{Cross-section of the process with the nuclear  $\gamma$ radiation }

We demonstrate our results on example of the $\eta_c(2.979)$ resonance
production.  The previous results (\cite{Baur}) used old values of its
widths and a point nuclear charge. Now we take resonance parameters from a new
Particle Date Group (\cite{PDG}) $\Gamma_{\eta_c\to\gamma \gamma} =
4.8$ keV and the realistic charge distribution.  
The calculations was made with the help of TPHIC event 
generator \cite{Hencken}.
We use a well known formula
\cite{Baur98} of a narrow resonance cross-section
\begin{equation}
  \label{49}
  \sigma_{\gamma\gamma \to X}(w_1,w_2) =
  8\pi^2(2\lambda_X+1)\Gamma_{X\to\gamma\gamma} \delta (W^2-M_X^2)/M_X
\end{equation}
where $W^2=4w_1w_2$, $\lambda_X$ and $M_X$ is a spin and a mass of the
resonance. The LHC luminosity and our results according to (\ref{33})
and (\ref{aaspectr}) are in Tab.~\ref{tab:xsec} for the process
(\ref{1}) with $A_{final}=A_1 $ or $A_1^*$.
\begin{table}[ht]
  \caption{Cross-section of {\mbox{\boldmath{$\eta_c$}}} 
    production by {\mbox{\boldmath{$\gamma \gamma$}}} fusion}
  \label{tab:xsec}
  \medskip
  \centerline{
    \begin{tabular}{lllccc}
      \hline
      $A_{final}$ & $L$ (cm$^{-2}$ s$^{-1}$) & $L$ (pb$^{-1}$) & $\sigma $ & 
      event/$10^6$ s &  \\
      \hline 
      \hline 
      \multicolumn{6}{c}{$\eta_c$ (2.979 GeV)} \\
      \hline
      \multicolumn{6}{c}{A point charge of the nuclei} \\
      \hline
      $^{208}{\mbox{Pb}}_{82}$ & $4.2\cdotp 10^{26}$ & $0.013$     & 356 $\mu$b       & $147000$ &  \\
      $^{~16}{\mbox O}_{8}$    & $1.4\cdotp 10^{31}$ & $441.5$     & 73 nb    & $1020000$ &  \\
      \hline
      \hline 
      \multicolumn{6}{c}{With form-factors of nucleus and in the region $R<b<\infty$} \\
      \hline
      $^{208}{\mbox{Pb}}_{82}$ & $4.2\cdotp 10^{26}$ & $0.013$     & 296 $\mu$b       & $122000$ &  \\
      $^{208}{\mbox{Pb}}_{82}^*~(3^-)$ & $4.2\cdotp 10^{26}$ & $0.013$     & 129 nb       & $53$ &  \\
      \hline
      $^{~16}{\mbox O}_{8}$   & $1.4\cdotp 10^{31}$ & $441.5$     & 66 nb     & $926000$ &  \\
      $^{~16}{\mbox O}_{8}^*~(2^+)$   & $1.4\cdotp 10^{31}$ & $441.5$     & 0.201 nb      & $~2810$ &  \\
      \hline
    \end{tabular}
  }
\end{table}

Our results from a Tab.~\ref{tab:xsec} shows that though the
cross-section of the process (\ref{1}) for the nucleus $^{208}$Pb is
larger than that for $^{16}$O, the event rate is smaller because of
the lower LHC luminosity for $^{208}$Pb.  The cross section with a
nuclear excitation is less by three orders of magnitude than that
without the excitation since the intensity of excitation is not large
and the inelastic form factor is less than the elastic form factor
(see Fig.\ref{fig1_gam} and Fig.\ref{fig2_gam}). Therefore for the
accepted LHC luminosities it is possible to use the secondary photons
as a signature of the clear electromagnetic nuclear process only for
the production $X_f$ with rather large cross-section
$\sigma_{\gamma\gamma \to X}$. The light ions are more preferable than
the heavy ions to detect the nuclear $\gamma$ radiation.

\section{Conclusion }

 In the work we suggest a new signature of the peripheral
ion collisions.

The formalism of the process (\ref{1}) is developed in the frame of
the equivalent photon approximation. New point is an introduction of
the inelastic nuclear form factor. It allows to consider the
excitation of discrete nuclear levels and their following $\gamma$
radiation decay.  It is shown that the energy of this secondary
photons are in GeV region due to a large Lorentz boost at LHC energies.
The angular distribution of the photons has a peculiar form as a function
of polar angle in the beam direction. The most photons hit the
region of ZDC in CMS and ALICE experiments in the region of angles of
a few hundred micro-radians.

So the nuclear $\gamma$ radiation is a good signature of the clear
peri\-pheral ion collisions at LHC energies when $A$ and $Z$ of beam ion
are conserved. The trigger requirements will include a signal in the central
rapidity region of particles from $X_f$ decay, a signal of photons in
the electromagnetic detector in front of the zero degree calorimeter
and a veto signal of neutrons in ZDC. We suggest to use the veto
signal of neutron in order to avoid the processes with nuclear decay
into nucleon fragments. The nuclear $\gamma$ radiation can be used for 
tagging the events with particle production
in the central rapidity region in the ultra-peripheral collisions.

The light nuclei are more preferable comparing with heavy ions since
they have higher beam luminosity at LHC.  The cross-sections of the
process with the nuclear excitation is three orders of magnitude
smaller than one without excitation.  The accepted nuclear
luminosities enable to use this signature for the large cross section
of $X_f$ system production.

Authors are very grateful to L.I.Sarycheva and S.A.Sadovsky
for the useful discussions.  \vskip 5mm


\end{large}
\end{document}